# Glass transition of an epoxy resin induced by temperature, pressure and chemical conversion: a configurational entropy rationale

Silvia Corezzi

*INFM and Dipartimento di Fisica, Università di Perugia, via Pascoli I-06123, Perugia (Italy)*

e-mail: Silvia.Corezzi@fisica.unipg.it

ABSTRACT

A comparative study is reported on the dynamics of a glass-forming epoxy resin when the glass transition is approached through different paths: cooling, compression, and polymerization. In particular, the influence of temperature, pressure and chemical conversion on the dynamics has been investigated by dielectric spectroscopy. Deep similarities are found in dynamic properties. A unified reading of our experimental results for the structural relaxation time is given in the framework of the Adam-Gibbs theory. The quantitative agreement with the experimental data is remarkable, joined with physical values of the fitting parameters. In particular, the fitting function of the isothermal $\tau(P)$ data gives a well reasonable prediction for the molar thermal expansion of the neat system, and the fitting function of the isobaric-isothermal $\tau(C)$ data under step-polymerization conforms to the prediction of diverging $\tau$ at complete conversion of the system.

## INTRODUCTION

As well-known one of the most prominent sign (let's say the fingerprint) of approaching the glass transition (GT) is the huge increase of the structural relaxation time, while the system remains in an amorphous phase. It is still known that the temperature is the variable most frequently used to drive a glass-forming system toward the glass transition. However, the use of pressure as an additional thermodynamic variable can provide an alternative route to the glassy state, being the effect on molecular motions of an isothermal increase of pressure quite similar to that produced by an isobaric decrease of temperature. Beside this, polymeric systems provide a further interesting route to the glass transition, through the increase of connectivity as the result of a polymerization reaction. Starting from a liquid system (with typical relaxation times of the order of picoseconds), one can get an amorphous solid (with typical relaxation times longer than hundred seconds), by changing both thermodynamic variables, like temperature and pressure (in which case one speaks about physical vitrification process), or a chemical variable, like the number of covalent bonds formed during a polymerization reaction (in which case one speaks about chemical vitrification process). The natural question that one faces is : To what extent is it possible to get an unified description of the glass transition phenomenon from a comparative analysis of these different ways to entry the glassy state ?

From the dielectric response of an epoxy system following different vitrification paths, we will appreciate the existence of some differences which are rationalized in terms of the chemical changes occurring in the reacting system. More interestingly, from similarities in dielectric relaxation times we will give an unified description of the dynamics close to the glass transition in terms of changes in the configurational entropy of the systems.

## EXPERIMENTAL AND DISCUSSION

In this work we considered a commercial grade sample of an epoxy resin, diglycidyl-ether of bisphenol-A (DGEBA), obtained by Shell co. under the trade name of Epon828 ($M_w \approx 380$) , and we studied its dynamics by wideband dielectric spectroscopy. The choice of this compound was guided by different reasons:

a) The basic one is that we can achieve the formation of a glass starting from the same ingredient, Epon828, but along three different paths. In fact, the neat resin is a very good glass-former, easily avoiding the crystallisation under cooling and compression. Moreover, it enters as a reagent in many polymerization reactions; in particular, if reacted with a bi-functional amine, like butylamine (BAM), yields a linear polymer through a linear step polymerization reaction, while if reacted with a poly-functional amine, like ethylenediamine (EDA), yields a network polymer, through a step reaction mechanism again.

b) Epon828 is a good candidate to be investigated by dielectric spectroscopy since it provides a strong dielectric response due to the reorientation of the strong dipole moment associated with the epoxy group, and also exhibits an appreciable conductivity due to the presence of ionic impurities.

c) Finally, but not the least important, it is a common epoxy resin for industry uses.

We reacted at 298 K the mixture Epon828/BAM in stoichiometric ratio 1:1, and the mixture Epon828/EDA in non-stoichiometric ratio 1:1. Moreover, the neat resin was measured at ambient pressure in the temperature range 163-363 K, and at room temperature in the pressure range 0.1-235 MPa. More experimental details are elsewhere reported [1,2,3].

Some common features can be recognized in the behavior of the dielectric response of the system. In all cases we have two dielectric processes: a secondary process at higher frequencies and a structural process al lower frequencies. The structural peak fast shifts toward lower frequencies, and the molecular motions responsible for the process slow down, either by decreasing the temperature, or by increasing the pressure, or by advancing the chemical reaction, and in any case it conventionally defines the glass transition when attains the value of $10^2$ s. A pronounced decrease of the conductivity contribution accompanies the increase of the structural relaxation time. On the contrary, the secondary process appears to be far less affected by this phenomenon, and it shifts very slowly to lower frequency. Below the glass transition, only this secondary relaxation is active, typically very small and spreading over many frequency decades.

We fitted the complex dielectric permittivity, $\varepsilon=\varepsilon'-i\varepsilon''$, by the superposition of two Havriliak-Negami (HN) functions, accounting for the presence of two relaxation processes, plus a conductivity term $-i\sigma/\omega\varepsilon_V$ ($\varepsilon_v$ is the vacuum dielectric permittivity), accounting for the ionic drift:

$$\varepsilon(\omega)-\varepsilon_\infty=(\varepsilon_0-\varepsilon_1)L_1(\omega)+(\varepsilon_1-\varepsilon_\infty)L_2(\omega)-i\sigma/\omega\varepsilon_0 \; ; \qquad (1)$$

$\varepsilon_0$ and $\varepsilon_\infty$ represent the low and high frequency limits of the complex dielectric permittivity; $\varepsilon_0-\varepsilon_1$ and $\varepsilon_1-\varepsilon_\infty$ are the structural and secondary dielectric strengths; $L(\omega)=[1+(i\omega\tau)^m]^{-n/m}$ is the normalized HN function, with $\tau$ the relaxation time, $0\leq m\leq 1$ and $0\leq n\leq 1$ the shape parameters which characterise the power-law behaviour of $\varepsilon''$ in the low and high frequency limits respectively, and the indexes 1 and 2 refer to the structural and secondary relaxation, respectively. The fitting procedure, carried out on both $\varepsilon'$ and $\varepsilon''$ simultaneously, was adopted to describe the spectra when the temperature or the pressure was changed or at different times during the polymerization reactions.

A deeper analysis of the spectra shows the existence of both differences and analogies between the different paths to the GT.

Obviously, a very deep difference exists between a pure system and a curing system. In the pure system the molecular structure does not change by changing the temperature or pressure, and we have constant dipole moments as well as constant dipoles concentrations, apart from the effect due to a change in the mass density. Differently, a polymerisation reaction not only results in more or less complicated links between molecules which produce a change in the density of the material, but usually also modifies the dipoles in the system: some dipoles may disappear, sometimes being replaced by molecular groups with a different dipole moment. As expected, the main differences between the different paths to the GT mirror this continuous modification of the chemical structure as the number of chemical bonds increases in the curing system. In particular, they are reflected in static properties, such as $\varepsilon_0$, and structural properties, such as the shape parameters m, n.

*Behaviour of $\varepsilon_0$ and monitoring of conversion.* In Fig. 1 the change of $\varepsilon_0$ is shown during the different experiments: $\varepsilon_0$ increases by decreasing the temperature, as well as by increasing the pressure, although in a less pronounced manner. The trend is opposite for the linear and crosslinking reactions. According to the Kirkwood-Frolich-Onsager equation [4], the change in $\varepsilon_0$ is a combined effect of a change in temperature, in the mean square dipole moment, in the concentration of dipoles, and in $\varepsilon_\infty$. During an isothermal reaction, if the density increase alone is taken into account, $\varepsilon_0$ should increase rather than decrease. The observed trend must be attributed to the modification of dipoles associated to the formation of each chemical bond. In this way, we rationalize the decrease of $\varepsilon_0$ under polymerization as a consequence of the disappearance of dipoles in the growing macromolecular structure. This is demonstrated by the ability of this quantity to monitor the degree of chemical conversion of the reacting system. In fact, when (like in our cases) the mean square dipole moments of the molecular groups involved in the reaction remain constant, and only the dipole concentration is modified, then the change of the static dielectric permittivity is linearly dependent on the dipole concentration and, therefore, on the chemical conversion. In these conditions, the normalised value of the static permittivity, $C_{diel}(t)=[\varepsilon_0(t=0)-\varepsilon_0(t)]/[\varepsilon_0(t=0)-\varepsilon_0(t\rightarrow\infty)]$, closely parallels the calorimetric conversion (Fig. 2) and can be used to monitor the advancement of the reaction.

*Behaviour of the shape parameters.* Fig. 3, showing the shape parameters m and n for the structural relaxation versus the relaxation time, compares the shape of isochronal spectra acquired during the different experiments. By changing the physical variables, temperature and pressure, for any given relaxation time the relaxation shape is the same independently of the physical path followed. Moreover, the structural relaxation is almost shape-invariant. Differently, the reaction reduces the values of m and n, so that while the reaction proceeds and the molecular weight of the products increases, the structural relaxation develops into a broader and more asymmetrical process. This behaviour can be rationalized within models that relate m and n to inter- and intra-molecular interactions [5,6]: m correlates with long-range motions involving inter-molecular interactions; n correlates with local dynamics involving intra-molecular interactions. A possible interpretation of the results in Fig. 3 is that the molecular mechanism of relaxation is unchanged when physical variables are changed; on the other hand, the relaxation mechanism is similar in the linear and network polymerisations.

Conversely, relaxation times and conductivity are far less sensitive to details of the chemical structure, while they monitor the molecular diffusion. In particular, dielectric relaxation time probes the rotational diffusion of dipoles and conductivity probes the translational diffusion of ions.

In this connection, the main similarities between the different paths to the GT can be recognized in the following points: (a) the Debye-Stokes-Einstein (DSE) relation is fulfilled under cooling, compression, and polymerisation; (b) the structural relaxation time is well described by a Vogel-Fulcher (VF)-like equation, as a function of both temperature, pressure and conversion.

*DSE behaviour.* The DSE relation was formulated on the basis of a hydrodynamic model and gives a connection between the relaxation and the transport properties in a fluid. In the viscous regime, in a good approximation it can be given as $\sigma \cdot \tau \cong$ const, and simply states that the conductivity is expected to be in inverse proportion to the dielectric relaxation time.

In order to check this relation, Fig. 4 shows $\sigma$ against $\tau$ on a double-log scale. In this representation the DSE behaviour should correspond to a straight line with slope –1. The linear fits to the data, drawn with solid lines in Fig. 4, indicate that the prediction of the DSE equation is fulfilled by our systems not only in the temperature domain but along all the different paths to the GT, provided that during polymerisation the conductivity is weighed by the reciprocal average numerical degree of polymerization, $x_n(C)$. According to Deng and Martin [7], this last factor accounts for the difference between the diffusion coefficient of the ionic species from which $\sigma$ originates, and the diffusion coefficient of the growing macromolecular products to which the relaxing dipoles are bound, so that the product $\sigma x_n^{-1}$ is a sort of average diffusion coefficient of the growing macromolecular structure.

These findings confirm that the changes in both conductivity and structural relaxation time simply mirror the dramatic changes occurring in the molecular diffusion, as the diffusion is markedly affected by temperature, pressure, as well as by chemical conversion.

*Behaviour of the structural relaxation time.* The behavior of the relaxation times as a function of pressure (Fig. 5b) shows strong similarity with the behavior against the reciprocal temperature (Fig. 5a). In particular, the pressure dependence of the structural relaxation time can be described in a good approximation by a phenomenological function which parallels the VF equation with P in place of 1/T, $\tau=\tau_0\exp[B_P P/(P_0-P)]$. During polymerization, if conversion is used as independent variable, a VF-like function, $\tau=\tau_0\exp[B_C/(C_0-C)]$, can be still used to represent the increase of the structural relaxation time (Fig. 5c). It seems important to stress that a VF-like description can be given as a function of conversion, but in general cannot be given as a function of the reaction time. Indeed, at constant T and P, the changes occurring in a reacting system are controlled by the number of chemical bonds formed between monomers, not by the reaction time: In order to produce a modification of the system, it does not matter how long one has to wait but how many bonds are actually formed between molecules. Then we expect that the most suitable independent variable for monitoring the physical changes in the polymerizing system is the conversion $C(t)=N_t/N_0$ ($N_t$= functional groups that have reacted, $N_0$= functional groups at the beginning), that measures, at any time t, the advancement of the reaction through the fraction of functional groups that have actually reacted at that time.

The possibility to recognise a VF-like behaviour as a good approximation of the structural relaxation time for all the three routes to the glass transition can be regarded as an important sign of the existence of some common factor controlling the dynamics of glass forming systems close to the glass transition. This simple observation drives to believe that a theory which contains the basic ingredient for explaining the glass transition phenomenon should be able to give a simultaneous explanation of the dynamic behaviour observed under isobaric and isothermal conditions as well as under polymerization.

In this connection, the main claim of the present work is that an unified reading of our experimental results can be given within the framework of the Adam-Gibbs theory, in terms of changes in configurational entropy.

A theoretical explanation of the dynamic behaviour of glass forming systems close to the GT has been tried in various models, by invoking a number of different concepts. One of the most interesting interpretation was put forward by Adam and Gibbs (AG) [8], based on the concept of configurational entropy and the assumption of cooperatively rearranging regions (CRR).

Having the start from the observation that the sluggish relaxation behaviour governing the GT is a manifestation of a dearth of configurations, the theory explains the behaviour of the structural relaxation time in terms of the size of the CRR. The outcome of the theory is the following expression for the structural relaxation time:

$$t = t_0 \exp\left(\frac{z^* \Delta m}{k_B T}\right) \qquad \text{eq.(1)}$$

where z* is the smallest size of a cooperative region that can perform a rearrangement into another configuration, Δμ is a free-energy barrier hindering the rearrangement per molecule, and $k_B$ the Boltzmann's constant. The critical size z* is determined by configurational restrictions and then related to the molar configurational entropy of the system, $S_c$, by:

$$z^* = \frac{N_a s_c^*}{} \qquad \text{eq.(2)}$$

where $S_c$ is the configurational entropy, defined as $S_{melt}$-$S_{vibr}$ (i.e., the total entropy of the melt deprived of the vibrational contribution), and $s_c$* is a not specified critical value for the configurational entropy associated to one CRR. Combining eq.(1) and eq.(2) gives a simple link between τ and $S_c$:

$$t = t_0 \exp(\frac{s_c^* N_a \Delta m}{k_B T S_c}) = t_0 \exp(\frac{A}{T S_c}) \qquad \text{eq.(3)}$$

This theory has been one of the most successful in giving a rationale to the VF temperature dependence of τ. By writing $S_c$ as:

$$S_c(T) = \int_{T0}^{T} \frac{\Delta C_p(T')}{T'} dT' \qquad \text{eq.(4)}$$

with $S_c(T_0)=0$, and $\Delta C_p$ the difference in molar heat capacity between the melt and glass, it is enough to assume the simple form $\Delta C_p(T) \propto 1/T$, in order to reproduce a VF equation with the temperature where τ diverges at the temperature where the configurational entropy vanishes.

In the following, we will show that really this theory is able to reproduce not only the temperature dependence, but also the pressure and conversion dependence of the structural relaxation time data obtained in this study.

*Temperature dependence.* We have experimentally evaluated the configurational entropy $S_c(T)$ for Epon828 as the integral of the excess molar heat capacity according to eq.4, by using specific heat data obtained from temperature modulated DSC measurements [9]. As a remark, we notice that the quantity $\Delta C_p(T)$ being conceptually connected with the 'ideal glass', but such state not experimentally accessible, an experimental method for precisely determining $S_c$ is not available and one is forced to develop an approximate procedure for obtaining its value. Since a substantial part of the excess entropy of a glass over its crystal phase has recently been shown to be vibrational in origin [10], we have chosen to approximate $S_c \cong S^{melt} - S^{glass}$. The results for $S_c(T)$ are shown in inset of Fig.6.

Eq.3 predicts a linear relation between logτ and the reciprocal product $(S_c \cdot T)^{-1}$, as far as the weak temperature dependence of A can be neglected. A direct test of the AG formula for the structural relaxation time of Epon828 is given in Fig. 6. A linear relation of the data is fulfilled over a large temperature range. The onset of a deviation from the linear relation can be perceived at the highest temperatures. It is worth noting that for this check there is no need to approximate the data by a VF equation.

We simply conclude that the AG model is successful in describing the experimental data in the temperature range close to the glass transition.

*Pressure*. The AG formula (eq. 3) can be extended to include the effects of pressure, as recently shown by Casalini *et al.* [11] and Capaccioli *et al.* [12]. It is enough to write the pressure and temperature dependence of the configurational entropy by including together with the isobaric contribution related to the excess heat capacity the isothermal contribution related to the excess thermal expansion:

$$S_c(T,P) = S_c^{isobar} - S_c^{isoth} = \int_{T_0}^{T} \frac{\Delta C_p}{T'} dT' - \int_0^P \Delta\left(\frac{\partial V}{\partial T}\right)_{P'} dP' \qquad \text{eq.(5)}$$

The second integral in eq. (5) takes into account the reduction of the configurationale entropy expected as a result of an isothermal compression of the material. This term can be calculated using the pressure and temperature dependence of the volume of the melt as given by the Tait equation of state [13]:

$$V^m(T,P) = V^m(T,0)\left[1 - c\ln\left(1+\frac{P}{B(T)}\right)\right] \qquad \text{eq.(6)}$$

where $c$ is a dimensionless constant and $B(T)$ can be expressed by $B(T)=b_1\exp(-b_2T)$ [14,15], and by approximating the molar thermal expansivity of the glass, $(\partial V/\partial T)^{glass}$, as independent of pressure. As a result of these approximations, the isothermal reduction of the configurational entropy, $S_c^{isoth}$, is given by:

$$S_c^{isoth} = \left(\frac{\partial V^m}{\partial T}\right)_0 \left\{P + hcP - Bc\left(h+\frac{P}{B}\right)\ln\left(1+\frac{P}{B}\right)\right\} - P\left(\frac{\partial V}{\partial T}\right)^{glass} \qquad \text{eq.(7)}$$

where $h=1-b_2/\alpha$, with $\alpha$ the thermal expansion coefficient at atmospheric pressure. Substituting eq.7 into eq.3, together with the value of the isobaric contribution at atmospheric pressure $S_c^{293K}=56.5$ Jmol$^{-1}$K$^{-1}$ evaluated from the TMDSC calorimetric data, and the other parameters $(\partial V^m/\partial T)_0=2.08\cdot10\text{-}7$ m$^3$mol$^{-1}$K$^{-1}$, $B^{293K}=309$ MPa, h=-5.5, $c=9.1\cdot10^{-2}$ all evaluated from independent thermodynamic measurements, gives a function to fit the pressure dependent experimental data which contains only one free parameters, the molar thermal expansivity of the glass. In Fig.7 the solid line shows the result from the fit. The agreement with the experimental data is very good, and gives the value $(\partial V/\partial T)^{glass} = 6\cdot10^{-8}$ m$^3$mol$^{-1}$K$^{-1}$ for the molar thermal expansivity of the glass which is quite reasonable. It has also been shown [11,12] that the pressure extended AG model can be written in the form of a VF-like expression for $\tau$(T,P).

We simply conclude that the AG model is successful in describing the experimetnal data as a function of the pressure.

*Polymerization.* Concerning polymerization, it is useful to start from the AG formula in the form given by eq.1 instead of eq.3, and focus on z*, the size of the CRR. When a reaction grows a macromolecular structure, it seems reasonable that the number of molecular entities in one CRR grows in a manner similar to the number of monomers in a macromolecule. In the simplest way, we assume:

$$z^* \propto x_n \qquad \text{eq.(8)}$$

where $x_n$ is the average numerical degree of polymerization, defined as the average number of monomeric units in a polymeric molecule. In a step polymerisation reaction, $x_n$ is given as a function of conversion by [16,17]:

$$x_n(C) = \frac{1}{1-\frac{f}{2}C} \qquad \text{eq.(9)}$$

where $f$ is the average functionality of the system, defined as the average number of functional groups per monomer. In the reactions studied in this work, we have in both cases $f$=2, and then $x_n(C)=1/1-C$. Substituting eq. 9 into eq. 1, and assuming that the change of $\Delta\mu$ during the reaction can be neglected, is enough to reproduce a VF-like dependence on conversion:

$$\boldsymbol{t} = \boldsymbol{t}_0 \exp\left(\frac{\Delta m \frac{z^*}{x_n}}{k_B T(1-C)}\right) = \boldsymbol{t}_0 \exp\left(\frac{B_c}{1-C}\right) \qquad \text{eq.(10)}$$

with $B_c=\Delta\mu(z^*/x_n)(1/k_BT)=const$, and the value of conversion where the structural relaxation diverges predicted at C=1.

In Fig. 8 the solid lines represent the fits with a VF-like equation, $\tau=\tau_0\exp[B_C/(C_0-C)]$, for our step polymerizations. The fitting parameters are: $\log\tau_0^{-1}=10.0$, $B_c=2.8$, $C_0=1.03$ for the reaction Epon/BAM 1:1, and $\log\tau_0^{-1}=10.7$, $B_c=5.4$, $C_0=1.08$ for the reaction Epon/EDA 1:1. The agreement with the experimental data is remarkable, and the divergence of $\tau$ is found at $C\approx1$, according to the prediction.

Moreover, from a quantitative estimation using as a first approximation the value $s_c^*=k_B\ln2$ and $z^*\approx x_n$, we find for the free-energy barrier $\Delta\mu$ values of the order of 2-3 Kcal/mole, which are well in the range of the molecular interaction energies, and for the number of monomers in one CRR at the GT values of the order of 5-10.

We simply conclude that the AG model is also successful in describing our experimental $\tau$ data as a function of conversion.

## CONCLUSIONS

In this work we have shown, by a dielectric investigation of an epoxy resin, how the glass transition can be driven by physical processes (cooling, compression), and chemical processes (polymerization).

The differences between the change of dielectric properties along the three different paths to the GT mirror the continuous modification of the chemical structure.

The similarities concern dynamic properties and mainly mirror the fact that structural relaxation time and conductivity probe the molecular diffusion.

Concerning the starting question: “To what extent is it possible an unified description of the GT phenomenon from a comparative analysis?”, we have shown that the AG theory of configurational entropy is able to reproduce in essence not only the temperature, but also the pressure and conversion dependence of the structural relaxation time data obtained in this study. In this connection, it is worth noting the remarkable agreement with the experimental data, in spite of the many approximations involved in the original formulas, eq.1 and eq.3, as well as in their extensions to pressure and conversion. How general are these findings has to be tested on more different systems, and this study may open to further experimental and theoretical work.

The natural conclusion is that during an isothermal-isobaric polymerisation the growing number of covalent bonds in the reacting system plays a role that is quite similar to that of a temperature decrease or pressure increase: this role is to reduce the number of configurations available to the system, forcing the molecules to move cooperatively over an increasing length scale. Reducing configurational entropy makes a system to approach spontaneously the glassy state, and configurational restrictions can be imposed to the system either by reducing the thermal energy or the volume of the system, or by the formation of covalent bonds between the molecules. In this sense an increase of structural connectivity has the same effect on the molecular dynamics as cooling and compression.

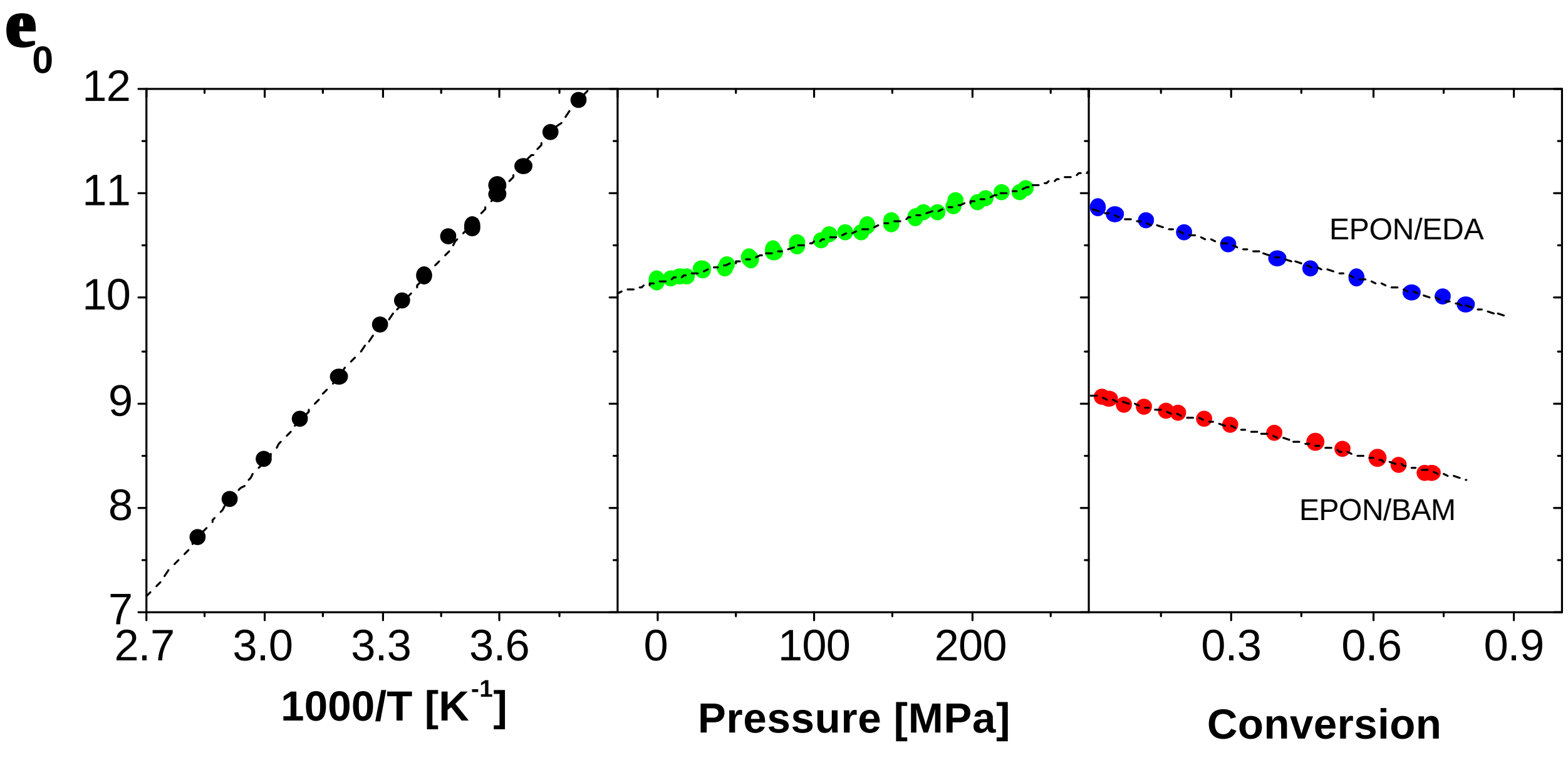

Fig. 1: Change of $\varepsilon_0$ during the different experiments: isobaric cooling at P=0.1 MPa, isothermal compression at T=293 K, and isothermal-isobaric polymerisations at T=298 K and P=0.1 MPa.

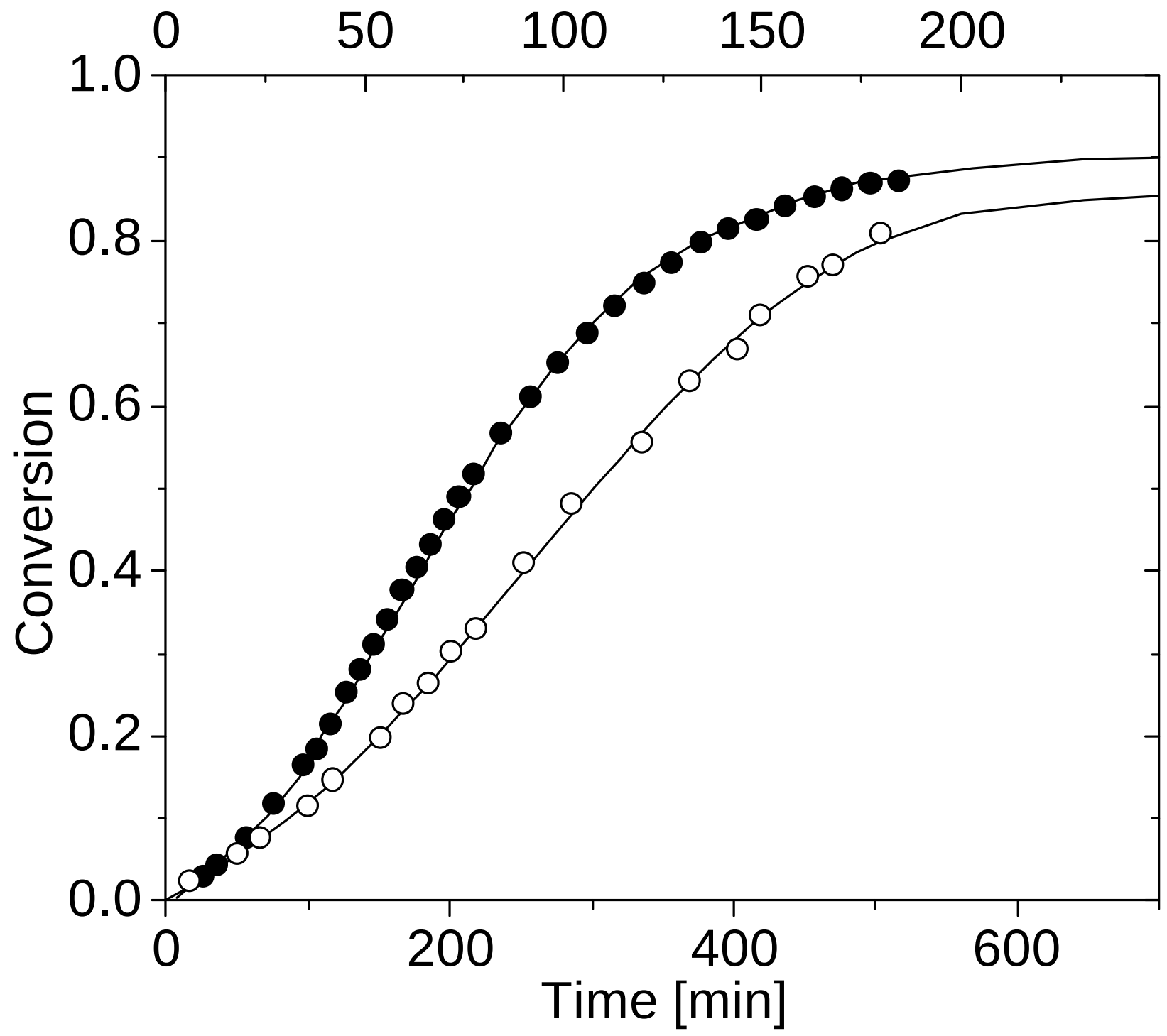

Fig. 2: Dielectric conversion, $C_{diel}$, of Epon828/BAM (solid circles; bottom axis), and Epon828/EDA (open circles; top axis). Solid lines represent the calorimetric conversion from DSC.

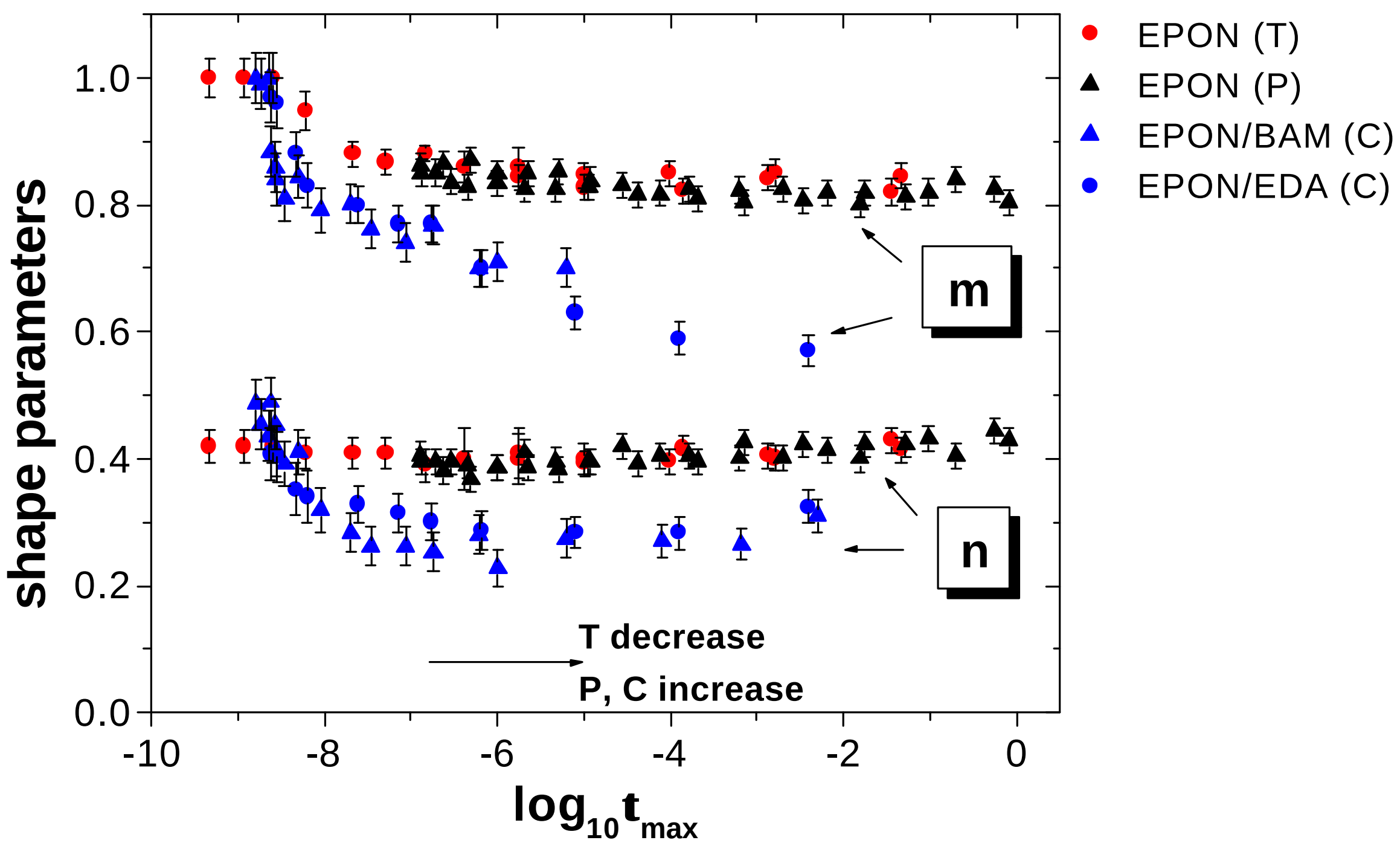

Fig. 3: Shape parameters, m and n, for the structural relaxation, versus the structural relaxation time. Red circles for the isobaric experiment at different temperatures, black triangles for the isothermal experiment at different pressures, blu triangles for the Epon828/BAM 1:1 reaction and blu circles for the Epon828/EDA reaction.

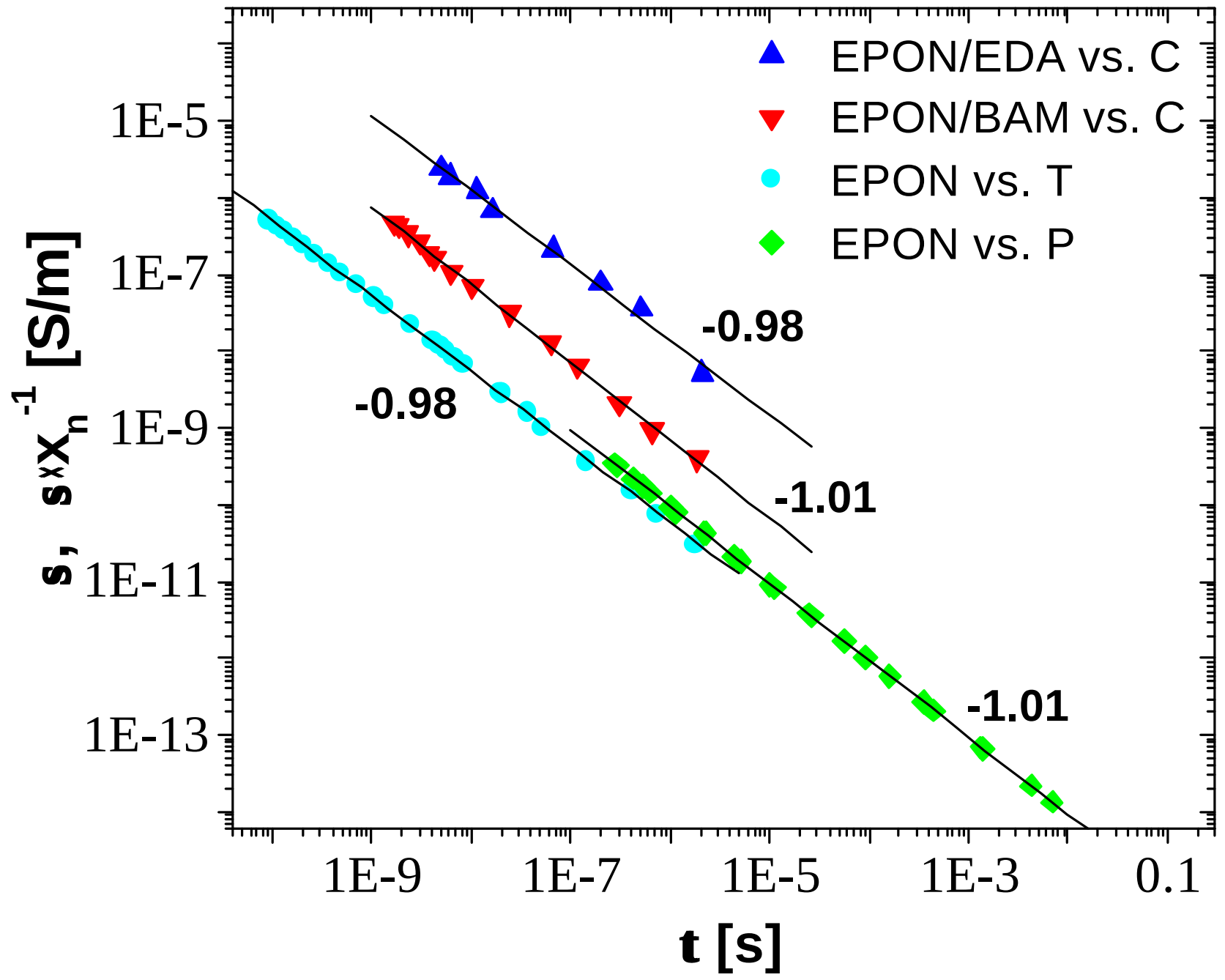

Fig. 4: σ vs. τ for Epon828 during the isobaric experiment (circles), and the isothermal experiment (diamonds). $\sigma \cdot x_n^{-1}$ vs. τ during the polymerization reactions with BAM (down triangles) and EDA (up triangles. The straigth lines represents the DSE relation.

**(a)**

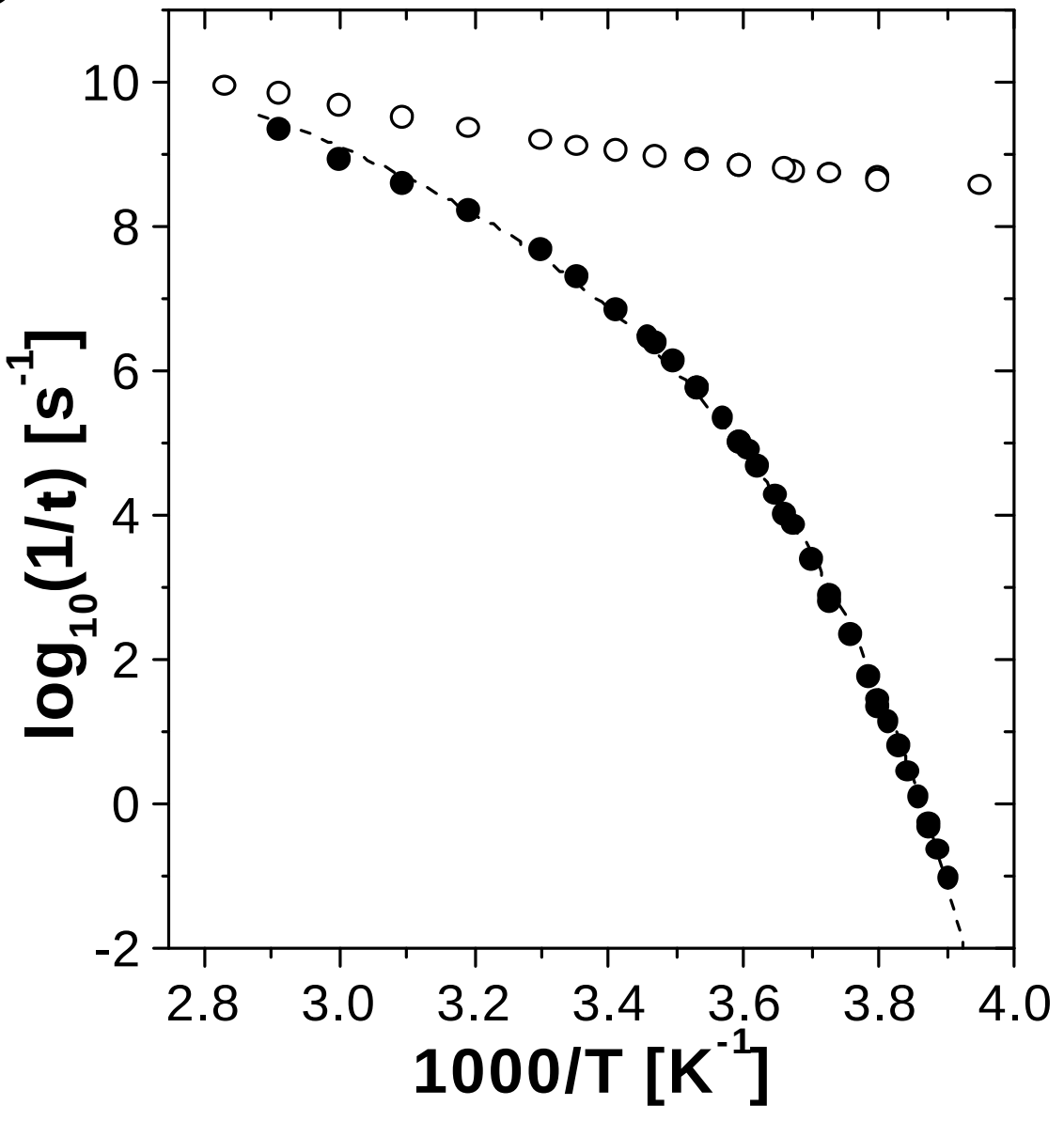

**(b)**

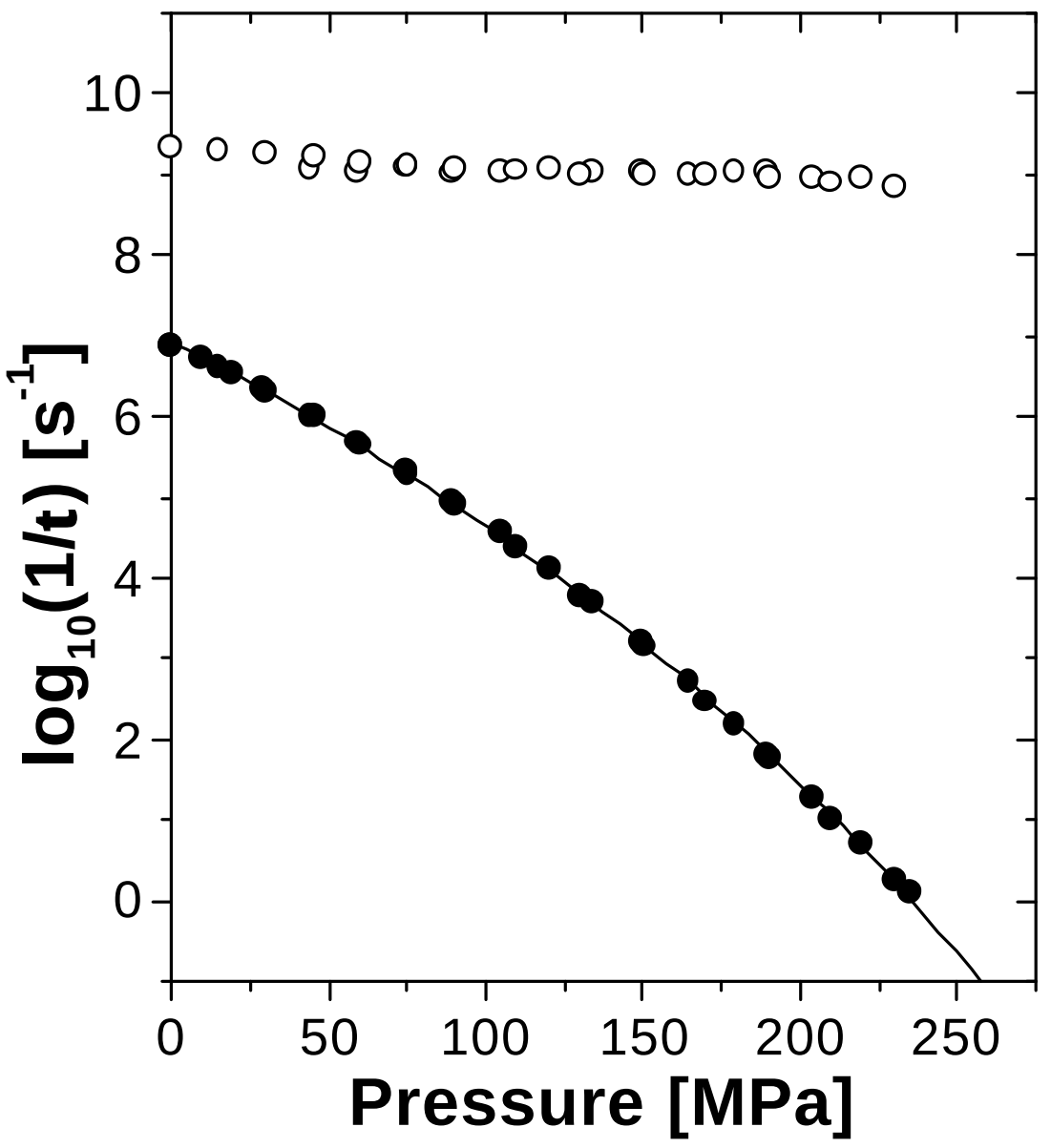

**(c)**

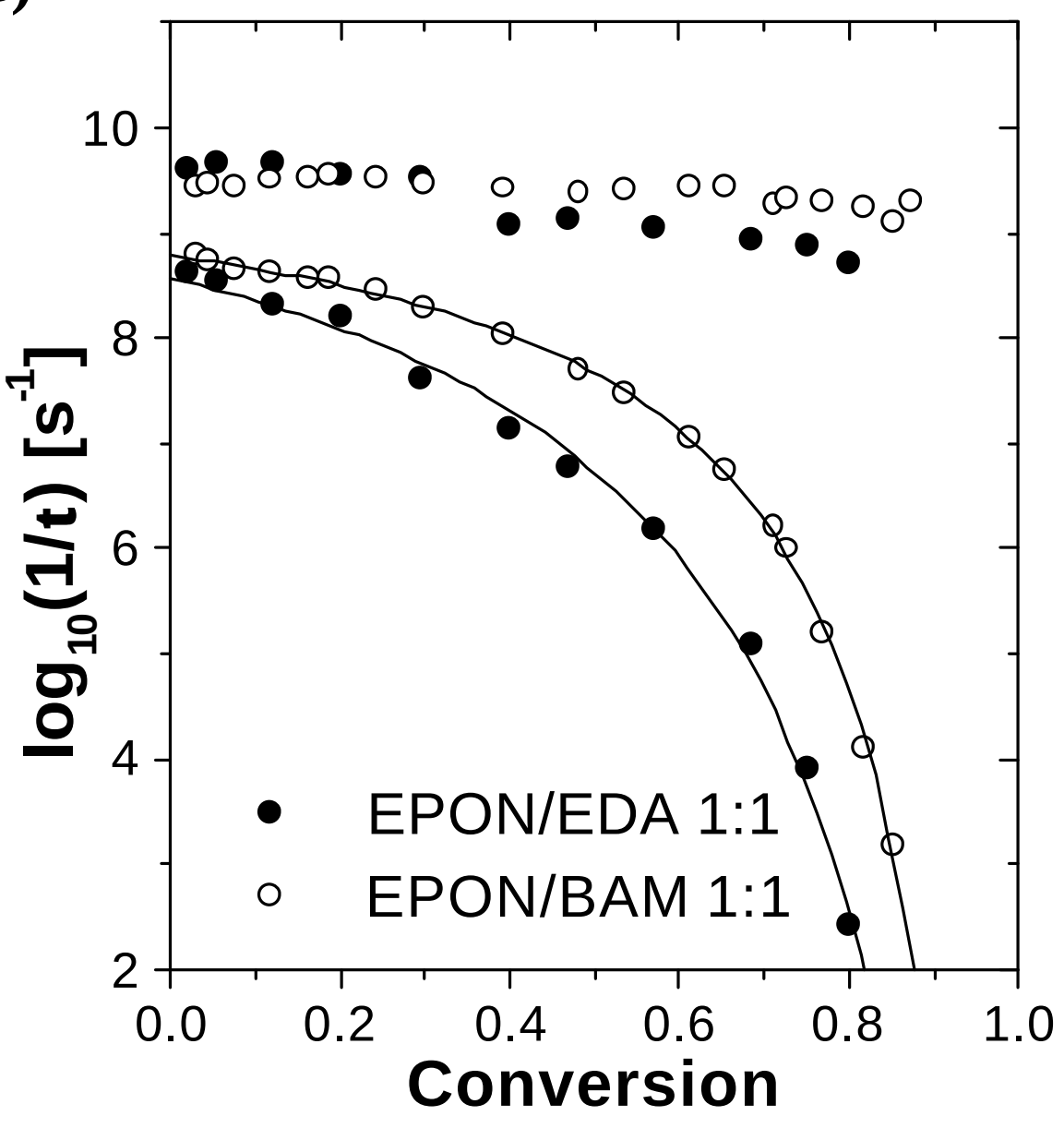

Fig. 5: Structural and secondary dielectric relaxation times during (a) the isobaric experiment, (b) the isothermal experiment, and (c) during the polymerization reactions with BAM and EDA. The lines are Vogel-Fulcher fitting curves.

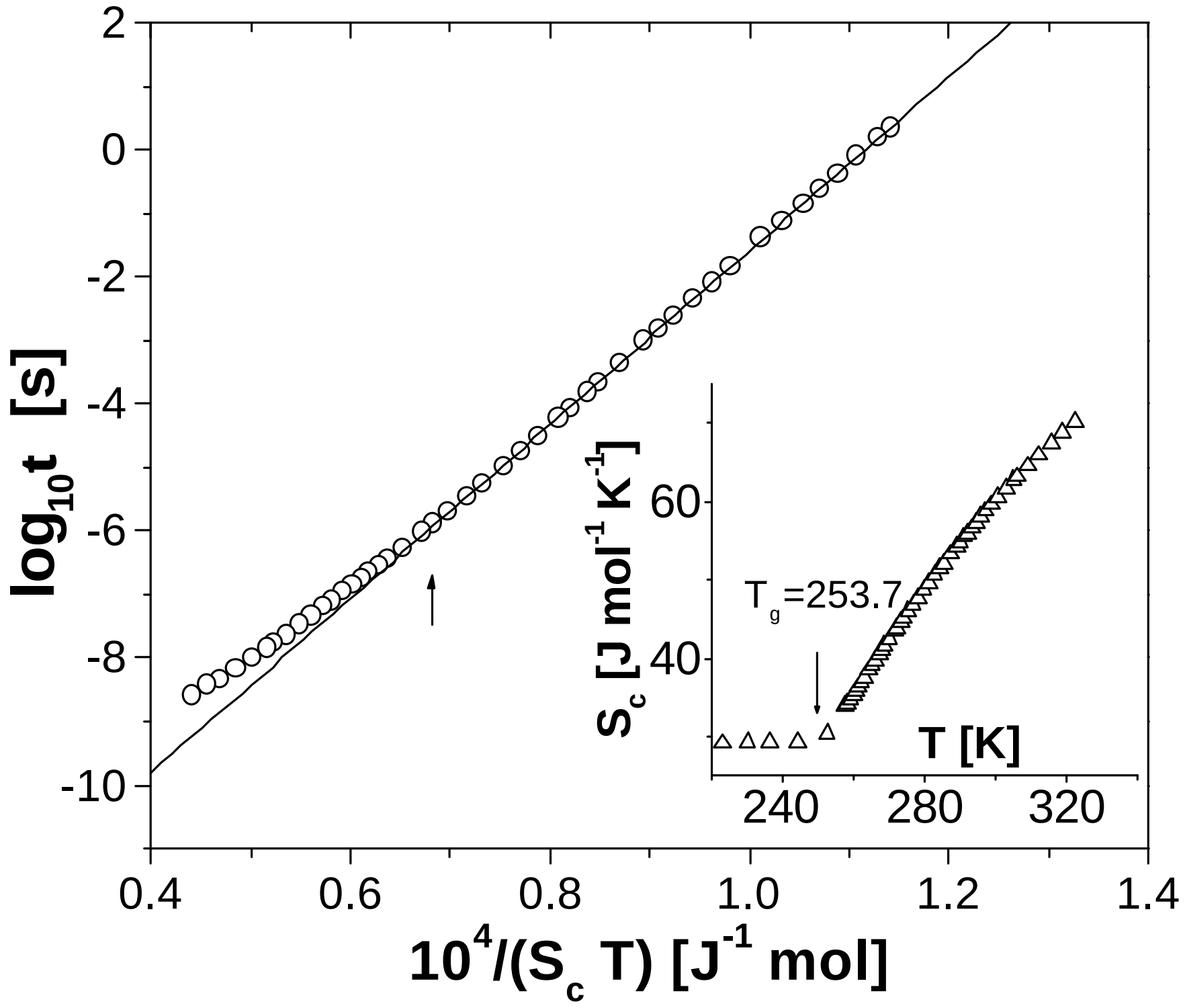

Fig. 6: $\mathrm{Log}_{10}\tau$ vs. $(S_c{\cdot}T)^{-1}$ for Epon828 at temperatures above $T_g$, with $\tau=1/2\pi\nu_{max}$ the dielectric relaxation time, and $S_c$ the molar configurational entropy evaluated as the excess entropy of the liquid over the glass. The solid line represents the Adam-Gibbs equation (3). The arrow indicates the onset of deviation from linearity. In the inset: temperature dependence of the configurational entropy.

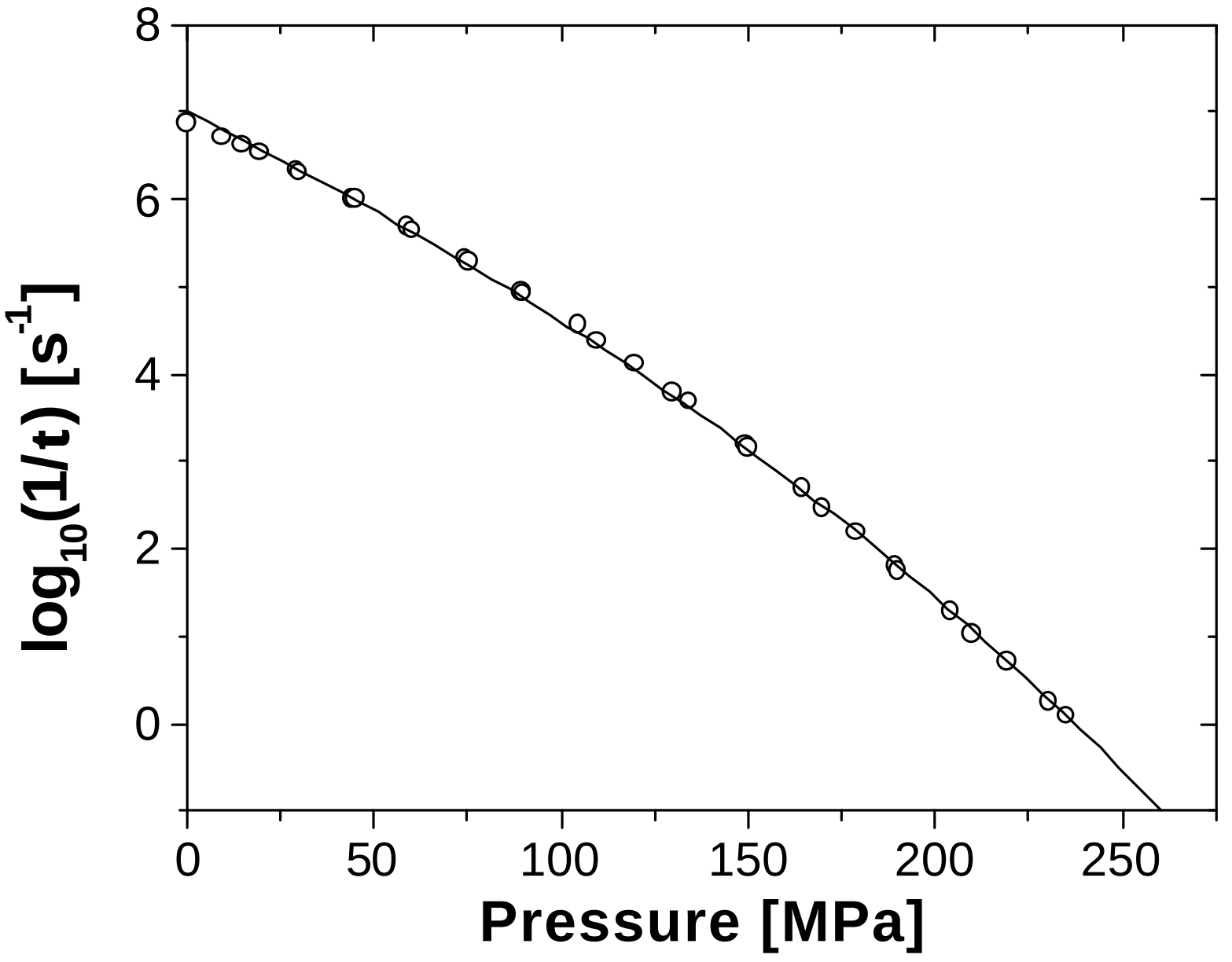

Fig. 7: Pressure dependence of the structural relaxation time of Epon828 during the isothermal experiment at T=293 K. The solid line is the fitting curve according to the pressure extended AG model.

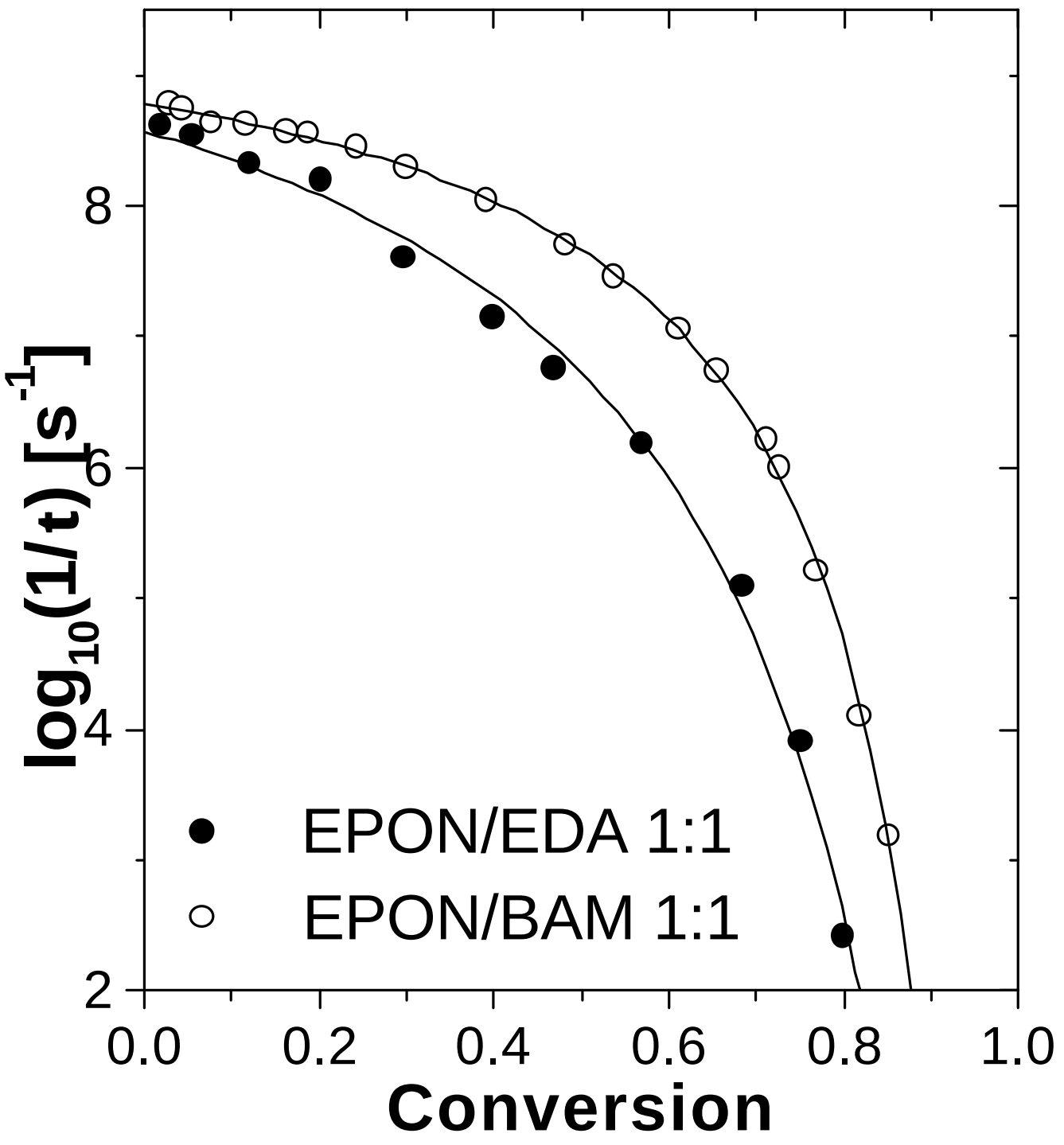

Fig. 8: Conversion dependence of the structural relaxation time of Epon828 during the linear step polymerization with BAM (open symbols) and the crosslinking step polymerization with EDA (full symbols). The solid lines represent the fitting curves with a VF-like equation, $\tau=\tau_0 \exp[B_C/(C_0-C)]$.